\documentclass[fleqn,10pt]{wlscirep}
\usepackage[utf8]{inputenc}
\usepackage[T1]{fontenc}
\usepackage{quotes}
\usepackage{sidecap}
\usepackage{comment}
\sidecaptionvpos{figure}{t} 

\title{Governance as a complex, networked, democratic, satisfiability problem}

\author[1,2,*]{Laurent H\'ebert-Dufresne}
\author[1,3]{Nicholas W. Landry}
\author[1,2]{Juniper Lovato}
\author[1]{Jonathan St-Onge}
\author[1,4]{Jean-Gabriel Young}
\author[5]{Marie-{\`E}ve Couture-M{\'e}nard}
\author[5]{St\'ephane Bernatchez}
\author[5]{Catherine Choquette}
\author[6]{Alan A. Cohen}
\affil[1]{Vermont Complex Systems Institute, University of Vermont, Burlington, VT, USA}
\affil[2]{Department of Computer Science, University of Vermont, Burlington, VT, USA}
\affil[3]{Department of Biology, University of Virginia, Charlottesville, VA, USA}
\affil[4]{Department of Mathematics and Statistics, University of Vermont, Burlington, VT, USA}
\affil[5]{Facult\'e de Droit, Université de Sherbrooke, Sherbrooke (Qu\'ebec), Canada}
\affil[6]{Butler Columbia Aging Center and Department of Environmental Health Sciences, Mailman School of Public Health, Columbia University, New York, NY, USA}

\affil[*]{laurent.hebert-dufresne@uvm.edu}

\begin{abstract}
Democratic governments comprise a subset of a population whose goal is to produce coherent decisions, solving societal challenges while respecting the will of the people.
New governance frameworks represent this as a social network rather than as a hierarchical pyramid with centralized authority.
But how should this network be structured?
We model the decisions a population must make as a satisfiability problem and the structure of information flow involved in decision-making as a social hypergraph.
This framework allows to consider different governance structures, from dictatorships to direct democracy.
Between these extremes, we find a regime of effective governance where small overlapping decision groups make specific decisions and share information.
Effective governance allows even incoherent or polarized populations to make coherent decisions at low coordination costs.
Beyond simulations, our conceptual framework can explore a wide range of governance strategies and their ability to tackle decision problems that challenge standard governments.
\end{abstract}
\begin{document}

\flushbottom
\maketitle

\thispagestyle{empty}

\section{Introduction}
\label{section:introduction}

Direct democracy might be an ideal for many, but it is burdensome and impractical to require every constituent of a population to vote on every single decision that must be made, or law that could be established.
Critically, decisions made in this way can often be contradictory as voters are neither fully rational \cite{wolfers2002voters} nor knowledgeable, and populations are often polarized \cite{smidt2017polarization}.
However, simplifying the governing structure too much can lead to a dictatorship or a small number of governing elites unlikely to represent the will of the entire population.
Therefore, we design different delegated governing models in order to reduce the coordination cost of governing and increase the coherence of decisions. 
But how should we structure this group of decision-makers? Who gets a say on what issue?

Different modeling frameworks have been developed to understand the emergence of various governing structures.
Drawing from dynamical systems, evolutionary theory, and complexity science, one can model the emergence and growth of centralized state authorities over time based on available resources and mechanisms such as resource transfers from the population to the governing authorities \cite{spencer1998mathematical}.
Or, drawing from a political science perspective, one can model the formation of governing structures as a game-theoretical problem that involves assumptions of office-seeking and policy-seeking or portfolio allocation between parties \cite{laver1998models}.
However, these different modeling frameworks all follow the paradigm of centralized authority, with a small number of elites or parties governing a large population.

\begin{SCfigure}
    \centering
    \includegraphics[width=0.7\linewidth]{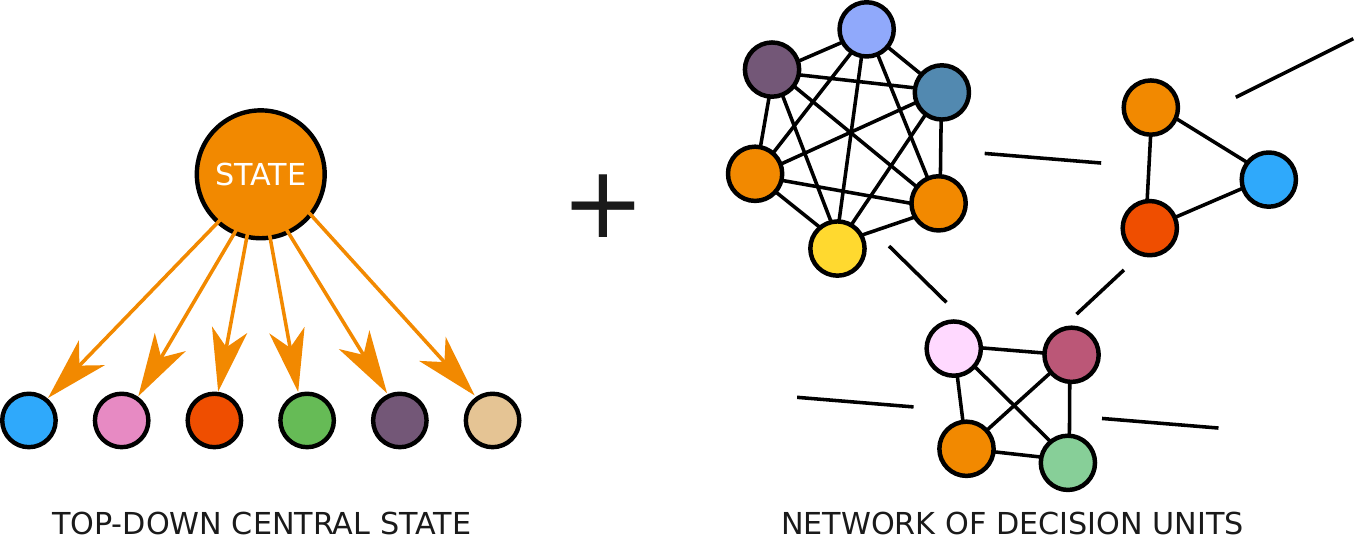}
    \caption{Governance: A top-down government structure supported by sectorial decision units\cite{choquette2021barrages}. Each decision unit represents a group made up of private stakeholders and government representatives. These decision units are each tasked with tackling a specific issue within their sector of interest and expertise.}
    \label{fig:choquette}
\end{SCfigure}

Over the last two decades, a specific model of social regulation has been emphasized.
This model, usually called governance, involves state and non-state members in decision-making processes.
The governance structure can be viewed as a social network rather than a hierarchical pyramid \cite{kerchove2019pyramide}.
It is also represented as a network of decision units, specialized in specific sectorial activities, which complete and support the traditional government structure of the State (see Fig.~\ref{fig:choquette}).

Governance is a paradigm shift argued by some to represent a new form of knowledge \cite{commaille2019metamorphoses} and by others to revisit ancient forms of social regulation \cite{rocher2019pluralite}.
In a network governance system\cite{keast2022network}, norms or administrative decisions are made by invested decision groups or stakeholders through consensus or vote.
Importantly, the participation of numerous people is expected to ensure more informed and largely accepted decisions, unlike other governing systems, where decisions are based on limited or biased knowledge or taken under undue pressure from lobbying groups\cite{fischer2012participatory}.
Governance efforts focus on stakeholders sharing information, ideally in a controlled democratic forum, and experiencing learning.
This participatory model of governance should result in stakeholders adopting effective behaviors, more adapted to societal crises that require quality data and changes of perspective, such as climate change \cite{choquette2021barrages} or COVID-19 \cite{couture2023expertise,couture2022gouvernance}.

In previous research, computational models of decentralized decision-making have embraced concepts of social influence \cite{axelrod1997dissemination} and self-organization\cite{rodriguez2007smartocracy} to capture the lack of a central decision maker.
Here, we propose a new framework to model such dynamics by more explicitly capturing the dynamic network structure of governance system.
Our framework is influenced by the modern practice of governance\cite{fischer2012participatory} and is designed to inform future applied efforts.

Network models of governance can take inspiration from, and conversely inform, our understanding of how other complex systems self-regulate; for example, how organisms maintain physiological homeostasis across diverse systems.
One of the most fundamental challenges an organism faces is to constantly and appropriately adjust countless aspects of its state in response to changing internal and external conditions \cite{cohen2012physiological,cohen2022complex, cohen2012physiological}.
In this sense, an organism or ecosystem could be considered analogous to a governance network system, organized at levels ranging from local to global and constantly making myriad decisions to ensure its functioning.
Many of the ``decisions'' made in biological systems ---- whether to experience hunger, how strong an immune response to mount, whether to go into cell cycle arrest, etc.---are informed by the integration of information about the biological state of the organism drawn from multiple other systems.
In turn, each of these factors may influence other decisions the organism must make, resulting in a coordinated suite of decisions being made constantly across the organism.
(Similar analogies might be made with other complex systems, such as ecosystems.)
Because biological systems have been fine-tuned by billions of years of natural selection, they have had the opportunity to refine efficient solutions to these kinds of regulatory problems.
Social regulatory systems have also evolved at their own pace, driven by the frequency of major crises and the rate of change in public perception.
Crises such as COVID-19 \cite{bernatchez2021liber}, or climate change and aging populations, may induce a shift toward network governance systems that are complementary to government systems.
This evolution involves a fine-tuning process, suggesting that there is ample room to improve governance models with thoughtful inspiration from other systems, as we attempt here.

In this paper, we develop a specific framework for the governance of complex systems as networks through higher-order interactions, which represent decision groups.
We first build a model of this networked governance, inspired by models from computational modeling and complex systems.
We explore only a small fraction of the space of potential scenarios to be modelled, but we show how our framework captures regimes of surprisingly effective governance.
We therefore discuss the potential use of this framework as an effective model, capturing both the coherence of centralized governments and the ideal of direct democracy.

\begin{figure}
    \centering
    \includegraphics[width=0.7\linewidth]{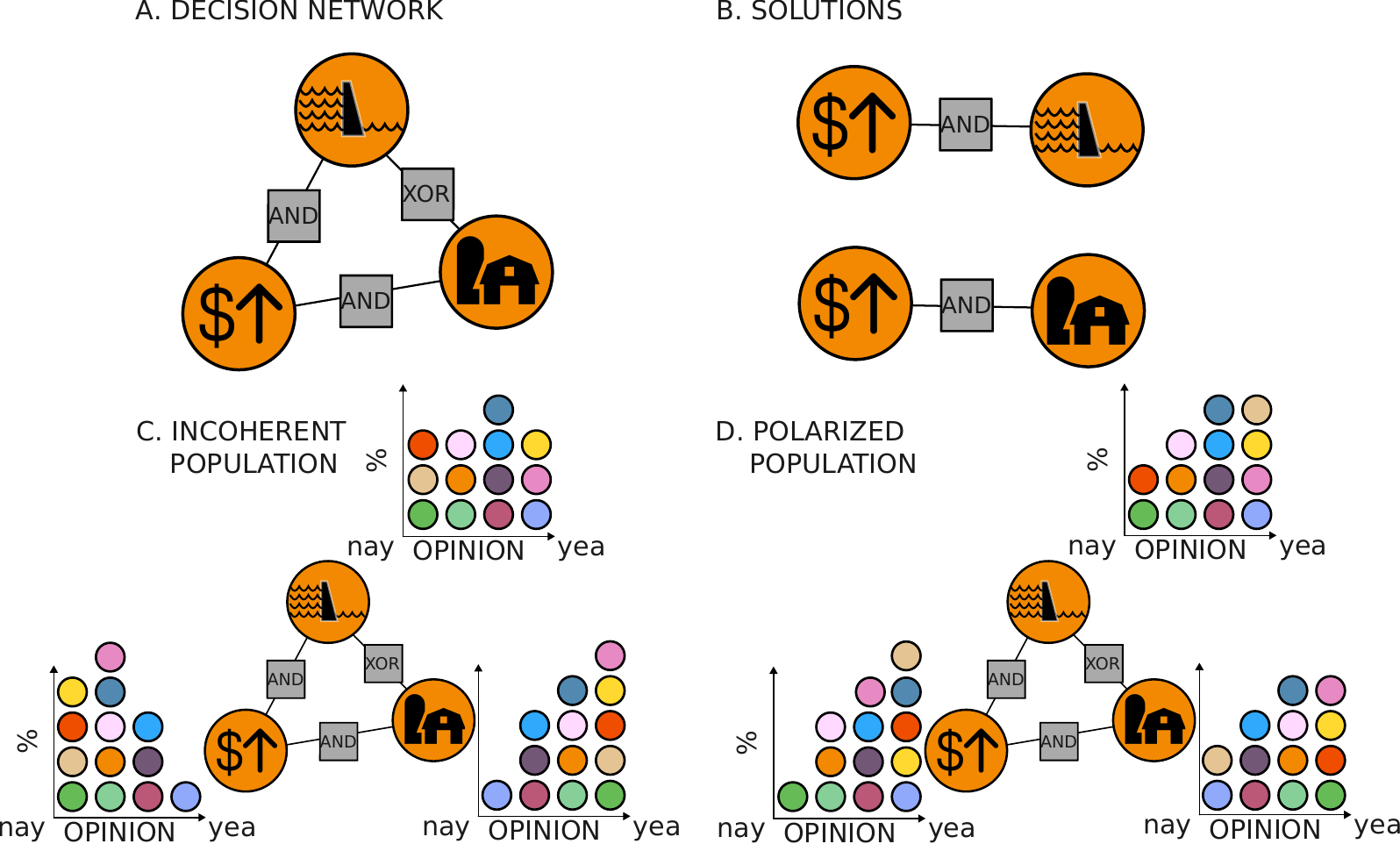}
    \caption{Schematic representation of our governance problem using signed networks with positive association denoted ``AND'' or negative association denoted ``XOR'' (disjunction/exclusive or).
    \textbf{A.} In this example, the decisions represent converting land into a hydroelectric reservoir (top), public farmland (right), and increasing local property taxes (left). Land conversion requires funding (``AND''), but the land can only be converted for a single use (``XOR'').
    \textbf{B.} Two simple solutions to the problem.
    \textbf{C.} We now represent the opinions of the 13 incoherent residents as a histogram. A direct vote would result in an incoherent set of decisions as the majority favors building a dam and farmland without raising taxes. The yellow agent is the most incoherent.
    \textbf{D.} A similar issue can arise in a coherent but polarized population, where one might end up in an incoherent state due to random fluctuations alone. Effective governance systems have to find the trade-offs and produce a coherent set of decisions that minimize democratic frustrations.}
    \label{fig:schematic}
\end{figure}

\section{Methods}
\label{section:model}

Our model combines two well-studied problems to capture the network effects involved in (1) logical constraints between decisions and (2) social consensus.
We call a ``decision'' any issue that requires an action in the form of a hard or soft legal norm: a law, recommendation, or guideline.
As a first approximation, we consider all decisions as binary: voting for or against a law, releasing a recommendation or not, etc.
Likewise, we consider the population as a large set of individuals fully determined by their own personal opinion about each decision, and thus devoid of personalities or dialectical abilities beyond quantifiable opinions.
The consensus of a group is therefore simply a function of the opinions of its members.
From these considerations, we split our governance model into two layers: the network of logical constraints between decisions and the governance network made up of groups tasked to make those decisions.

\subsection{Governance as a satisfiability problem}

First, we represent the logical constraints that arise in decision-making as a Boolean satisfiability problem (SAT).
Satisfiability problems have previously been used to model workflow execution, which can be compromised by the unavailability of specific agents, collaborators, or resources \cite{wang2007satisfiability}.
Such models have been used to study the adaptive nature of human systems during crises like COVID-19 \cite{mohammed2022pandemic}. 

In the SAT representation of governance, every decision can be resolved positively (voted for) or negatively (voted against), and to be coherent, decisions need to satisfy a logical structure.
For example, a governing body cannot simultaneously decide to convert land into a hydroelectric reservoir and into public farmland.
In SAT terms, the decisions are encoded as \emph{variables}, which can be either ``true'' (for) or `false'' (against), and logical constraints are encoded as clauses involving these variables, such as the AND clause (conjunction, ``all of''), the OR clause (disjunction, ``at least one of''), the XOR clause (exclusivity, ``only one of''),  and so on. 
Further, variables can be negated to express more nuanced sets of logical constraints.
By joining several such clauses together, one can express arbitrary logical requirements.

For example, suppose that something must be done about a vacant plot of land, and that it could be used as a reservoir (variable $x_1$) or public farmland (variable $x_2$), but not both.
Further, suppose that both land uses require increased property taxes to fund the conversion (variable $x_3$).
Any coherent set of decisions will need to satisfy the formula
\begin{equation*}
    (x_1 \lor x_2) \land (\lnot x_1 \land \lnot x_3) \land x_3
\end{equation*}
where $\lor$ denotes a logical, OR and $\land$ denotes a logical AND and $\lnot$ is a logical negation.
Thus, $(x_1,x_2,x_3)=(\mathrm{True}, \mathrm{False}, \mathrm{True})$ is an acceptable solution (the complete formula evaluates to ``True''), but $(x_1,x_2,x_3)=(\mathrm{True}, \mathrm{True}, \mathrm{True})$ is not.
Neither is inaction, $(x_1,x_2,x_3)=(\mathrm{False}, \mathrm{False}, \mathrm{False})$.

SAT problems like these can be represented as bipartite networks \cite{moore2011nature}, in which nodes correspond to variables and their negation, and clauses are presented by `factor nodes'' that are connected to the decisions involved in any given clause---see Fig.~\ref{fig:schematic}~A and B.
The logical constraints inherent to decision-making thus give rise to a first set of "network effects" in governance.

The SAT framing highlights two important aspects of governance.
First, deciding whether any valid truth assignment exists is NP-complete, while optimizing the number of satisfied clauses (say, to help reach social consensus) is NP-hard \cite{garey1979computers}.
As a result, if $\textsc{P}\neq \textsc{NP}$, we know that a population solving a complex decision problem \emph{has} to use shortcuts and simplifying heuristics.
In fact, not all networks are fully satisfiable, and the population will often be penalized for incoherent decisions, either due to the absence of a fully coherent solution or to the population's inability to find one. 
Second, these formal computational complexity results only concern the ``worst case''---i.e., algorithms designed to solve arbitrarily complex decision problems. 
Thankfully, under various generative models of how constraints arise, it is well known that typical or ``average case'' instances of the problem can be solved with simple methods and admits numerous viable solutions~\cite{moore2011nature}. 
Thus, we might hope that the typical challenges facing a society are not so hard---a fact that we will leverage in our experiment.

Next, we frame governance decisions as a \textit{democratic satisfiability problem} because we also want to consider the initial opinions of stakeholders and government agents within the population.
For simplicity, all of our experiments consist only of conjunction and exclusivity clauses.

\subsection{Governance as a hypergraph voter model}

Second, we represent decision-making within a community as a multidimensional hypergraph voter model \cite{hickok2022bounded}.
The voter model provides us with a simple discrete mechanism where nodes in the population (government actors and private stakeholders) update their state based on their neighbors in a growing social network structure. Here, this social network can be represented as a hypergraph consisting of decisional group interactions between agents involved in each governance unit for a specific issue.

A hypergraph, $H = (V, E)$ is a mathematical object that can encode group interactions between individuals in a population \cite{battiston2020networks}.
The set of nodes (or vertices), $V$, represents the entities in a system, while the set of hyperedges, $E$, is the collection of group interactions between those entities, with each interaction represented as a set.
An $m$-hyperedge $e$ in $E$ is a group interaction that involves $m$ entities, i.e., $e=\{i_1, \dots, i_m\}$, and when $m>2$, it is referred to as a \textit{higher-order interaction}.
In our model, nodes represent stakeholders and individuals in the population, while hyperedges represent groups of stakeholders deciding on a particular policy issue.
In contrast to the pairwise interactions, or \textit{edges}, that comprise graphs, hyperedges can overlap with one another in non-trivial ways \cite{aksoy2020hypernetwork}, via ``bridge'' nodes.
This overlapping structure provides a way to modulate the amount of information shared between groups of stakeholders through the number of ``bridge'' nodes.
Our higher-order social structure is not fixed from the start, but grows as policy issues are decided.
From that perspective, network growth has many parallels to choice theory based on the different ways in which we could grow the hypergraph in response to the opinion dynamics that unfold on it.\cite{overgoor2019choosing}.

A voter model is a binary-state agent-based model in which individuals adopt an opinion for or against an issue depending on the state of their neighborhood \cite{clifford1973model,holley1975ergodic,redner2019reality}.
The traditional version of this model considers a 2-dimensional lattice, where a node $i$ is selected uniformly at random and then randomly adopts the opinion of one of its neighbors, but other studies have considered more realistic structures and opinion adoption behavior.
Examples of more realistic network structures include small-worldness \cite{castellano2000nonequilibrium}, heterogeneous degree distributions \cite{sood2005voter}, and higher-order interactions \cite{horstmeyer2020adaptive}. 
Extensions of the opinion adoption mechanism include heterogeneous adoption rates \cite{masuda2010heterogeneous}, stubborn agents \cite{mobilia2007role}, consultation with multiple neighbors \cite{castellano2009nonlinear}, and more than two opinion states \cite{vazquez2004ultimate}.

In our model, nodes $i$ part of a hyperedge $e$ are tasked to make a certain decision $d_j$ on issue $j$, which can be positive ($d_j = +1$) or negative ($d_j = -1$).
Through discussion, the nodes $i$ observe the average opinion of the group $\langle x_j \rangle_e$ and update their opinion to $+1$ if this average is positive and to $-1$ if the average is negative.
We also assume that nodes discuss related issues $k$ that have logical constraints with the decision at hand ($k\in \mathcal{G}_j$, the neighbors of $j$ in $G$). If the decision $d_j$ is positive ($+1$) they update their opinion on $k$ to $+1$ if the constraint is conjunctive ($G_{j,k}$ = +1) and $-1$ if it is exclusive ($G_{j,k}$ = -1), and vice versa if the decision $d_j$ is negative.
Decision groups thus reach a binary and coherent solution to a simple decision problem centered on the decision they were tasked to tackle.

\begin{SCfigure}
    \centering
    \includegraphics[width=0.6\linewidth]{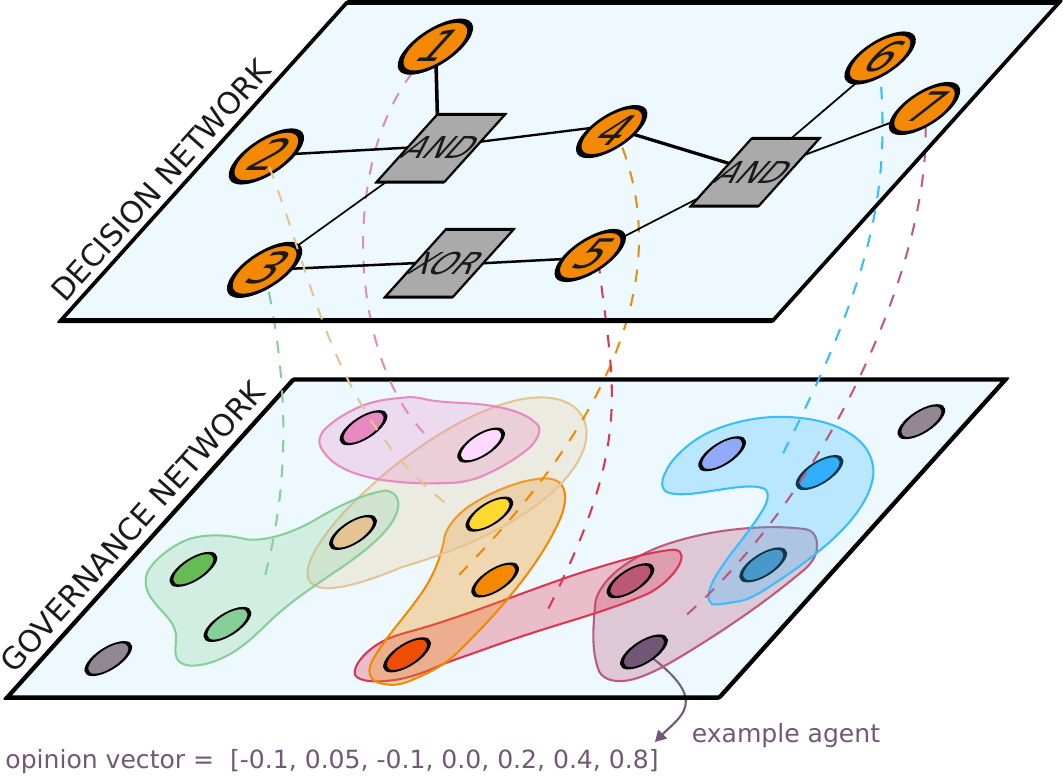}
    \caption{Schematic representation of our model.
    Here, a population of 15 agents needs to make 7 decisions.
    The constraints between these decisions are captured by a satisfiability problem shown in the top layer, i.e., the decision network, as described in Fig.~\ref{fig:schematic}.
    The opinion of each agent regarding each decision is captured by their individual opinion vector of length 7.
    Solving the decision network while respecting the democratic will of the population is a complex problem.
    To do so, the population forms a hypergraph structured through 7 groups, or hyperedges, represented as blobs that replace the cliques in Fig.~\ref{fig:choquette}.
    In each hyperedge, agents discuss their assigned decisions and related ones, reaching a consensus on these dimensions of their opinion vectors.
    This consensus is used to make a decision and attempt to solve the decision problem.}
    \label{fig:model}
\end{SCfigure}

\subsection{Summary of our model of governance}

Our theoretical model of governance is represented in Fig.~\ref{fig:model} and can be described as follows for a population of $N$ agents looking to make $D$ decisions under a network of constraints $G$.
\begin{enumerate}
    \item At each time step, a hyperedge $e$ of a given group size is formed to make a specific decision $d_j$ on issue $j \in G$ using network growth processes \cite{hebert2011structural} or random selection.
    We use random selection, where we fix the group size and overlap between the groups.
    Overlaps consist of randomly selected nodes that have previously made related decisions (if any).
    All other nodes are then selected uniformly at random to obtain the correct group size.
    This hyperedge $e$ is then added to the decision hypergraph, $H$.
    
    \item The nodes within group $e$ then update their opinion $x_j$ on $j$ and on issues neighboring $j$ in $G$.
    This reflects the nature of discussions within a group.
    Opinions about $j$ are updated to the average opinion of the group (or some other consensus function).
    
    \item Using their updated opinion, the group makes the decision $d_j$.
    The agents in the group also update their opinions on decisions $k$ in the neighborhood $G_j$ of $j$ in the decision network to ensure the consistency of their own opinions after discussions.
    For example, if a positive decision $d_j = +1$ was made on $j$, all agents would have an opinion of +1 on decisions in conjunction with $j$ ($G_{j,k} = +1$) or -1 on decisions exclusive to $j$ ($G_{j,k} = -1$).
    
    \item The process repeats until all decisions are made and outputs a set of decisions and a hypergraph $H$.
    
    \item We then evaluate the fraction $\mathcal{C}$ of all $\vert G\vert$ constraints in $G$ that are solved (coherence of the resulting decisions) and the satisfaction $\mathcal{S}$ of democratic choices (sum of the product of decisions and average initial opinion).
    The performance of the hypergraph $H$ in making a vector of decisions $d$ solving the constraint graph $G$ in a population $V$ of agents with opinions $x$ is summarized by these two scores:
    \begin{eqnarray}
        \mathcal{C} & = & \frac{1}{2\vert G\vert}\sum _{j\in G} \sum_{k\in G_j} d_j d_k G_{j,k} = \frac{1}{2\vert G\vert}d^\top G d\; ,\\
        \mathcal{S} & = & \sum _{j\in G}  d_j \langle x_j \rangle_V = \langle d^\top x \rangle_V  \; .
    \end{eqnarray}
\end{enumerate}

\section{Results}
\label{section:results}

\subsection{Simple problems of polarized or incoherent populations}

We first run our model on a trivial set of constraints that are easily satisfied.
We use a network of 6 decisions structured as a star network where four leaf nodes have a disjunctive constraint with the root node, and one leaf node has a conjunctive constraint with the root node.
The root node should therefore be aligned with the conjunctive leaf (i.e., both at +1 or -1) and misaligned with all other decisions.
The problem is simple, but since there exist two symmetric solutions, it can be made harder when it involves the wrong population.

\begin{figure}
    \centering
    \includegraphics[width=0.7\linewidth]{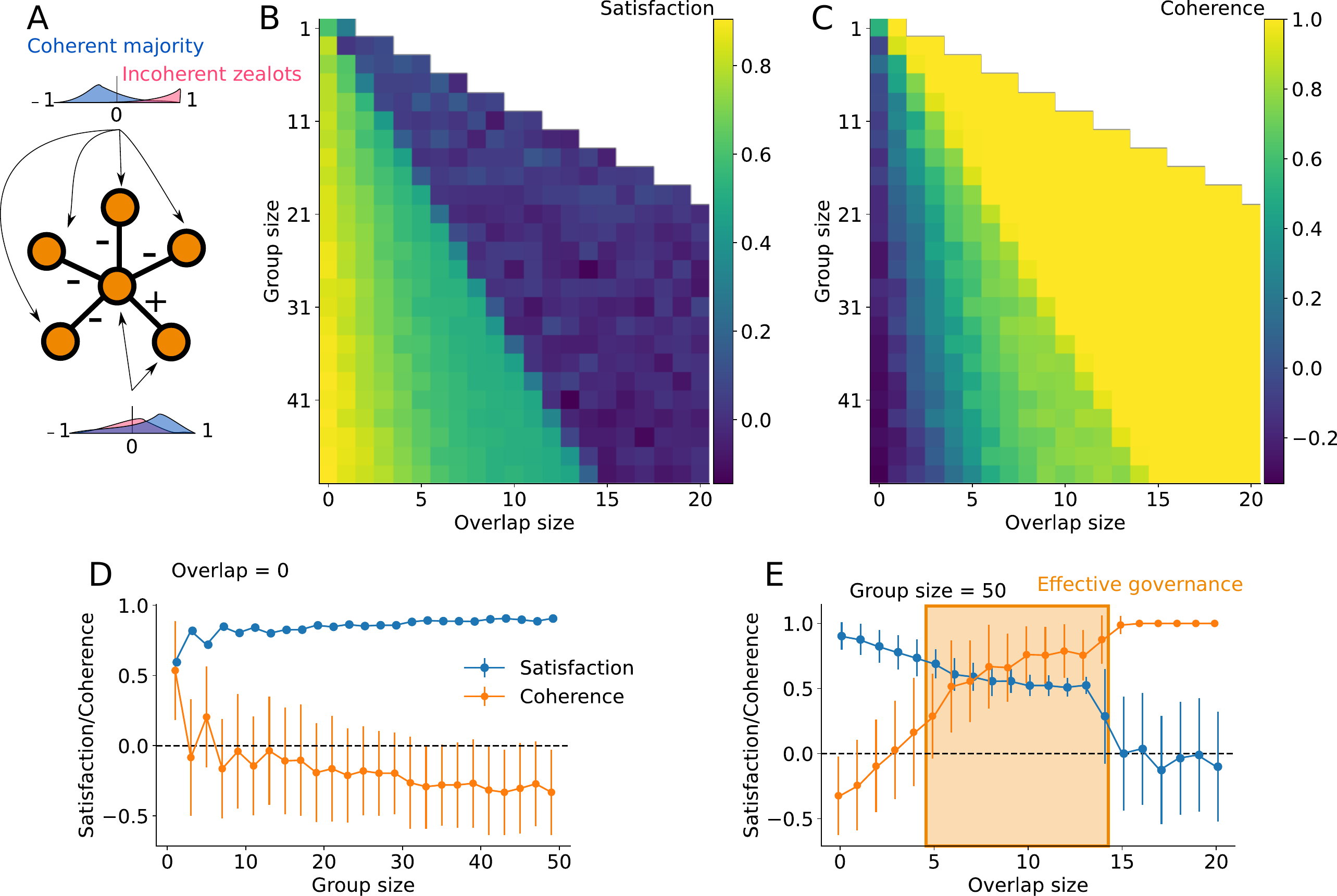}
    \caption{Governance model with a coherent majority and incoherent zealots. The simple decision network is shown in panel \textbf{(A)} and admits two complementary solutions.
    The population making the decision is described in the main text, and the distributions of their opinions, with a coherent majority and incoherent zealots, are shown in cartoon form. 
    In panels \textbf{(B\&C)}, we show the average democratic satisfaction and coherence of the decisions made by a given governance structure.
    Panels \textbf{(D\&E)} show two slices of the previous panels, where error bars represent the standard deviation of the governance outcomes.
    \textbf{(B\&D)} We see that increasing the size of the decision groups provides better statistical sampling of the population, increasing satisfaction but hurting coherence as more and more zealots get involved in decisions.
    \textbf{(C\&E)} Importantly, adding a small amount of overlap between decision groups can dramatically increase coherence, counteracting the role of zealots without hurting democratic satisfaction as much.
    This region, highlighted in the right panels, is what we call \textit{effective governance}, where solutions with both high satisfaction and effective governance are achieved.}
    \label{fig:zealots}
\end{figure}

Consider a population where a majority of agents are coherent with reasonable but mild opinions that solve a simple decision network, while a minority of "zealots" have incoherent and much stronger opinions about certain decisions.
We fix the total population to 10,000 agents with 30\% of zealots. Non-zealots have coherent opinions drawn from truncated normal distribution between -1 and 1, with standard deviation 0.1 and means equal to 0.43 for the two decisions in the positive (``and'') associative clause and -0.29 for all others. Zealots are focused on specific issues, and have opinions drawn from the same truncated normal but with mean opinions 0.28 and 1.0.
We then vary the group size and overlap between discussion groups to explore how a networked governance structure can help maximize both democratic satisfaction in the decisions made and the coherence of those decisions given the constraint network.
The results are shown in Fig.~\ref{fig:zealots}.

\begin{figure}
    \centering
    \includegraphics[width=\linewidth]{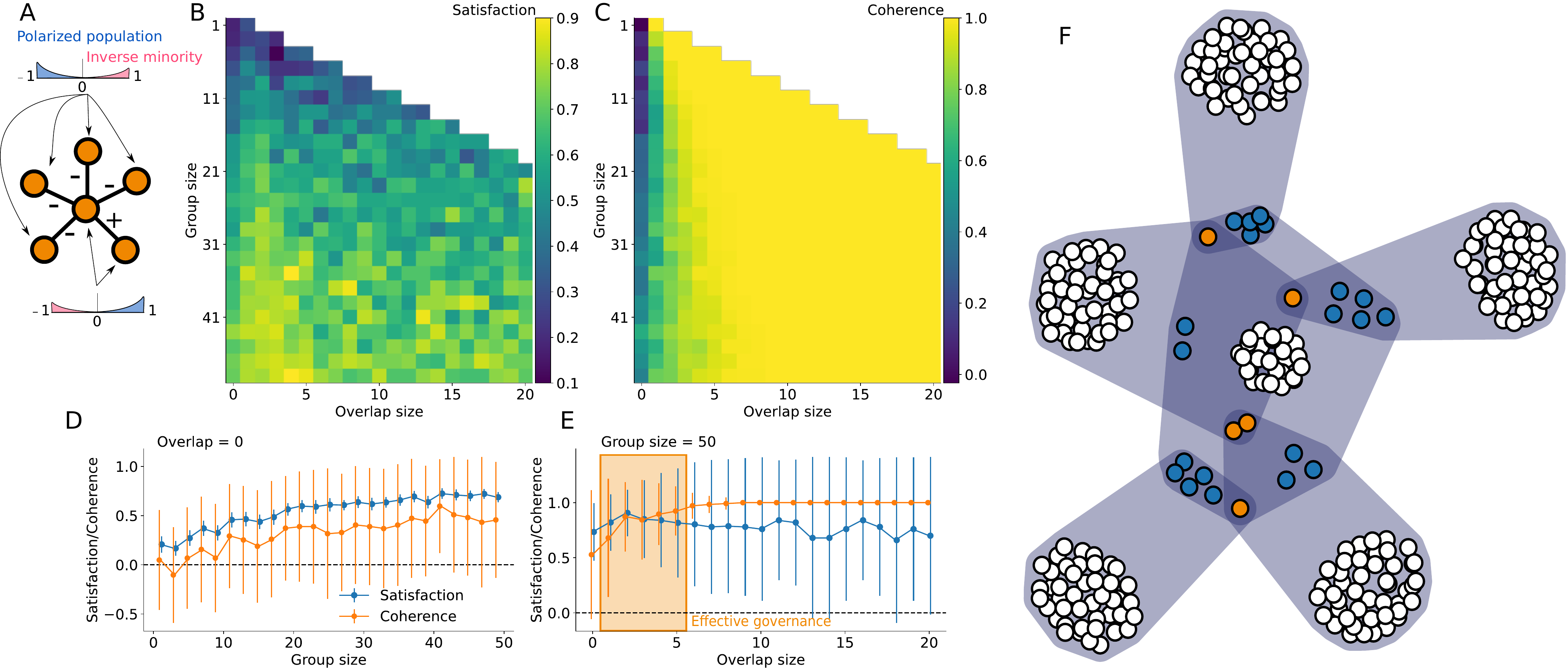}
    \caption{Governance model with a polarized population. The simple decision network is illustrated in panel \textbf{A} and admits two complementary solutions.
    The population making the decision is described in the main text and the distributions of their opinions, with asymmetric polarization, are shown in cartoon form.
    Increasing the size of the decision groups provides better statistical sampling of the population, increasing satisfaction (panel \textbf{B}) and coherence (panel \textbf{C}).
    Importantly, compared to a scenario with no overlap (panel \textbf{D}), adding a small amount of overlap between decision groups (panel \textbf{E}) can dramatically increase coherence and surprisingly increase satisfaction. Panel \textbf{F} shows an example of a hypergraph structure of decision groups produced around the \textit{effective governance} regime of panel \textbf{E} (group size of 51 with overlap of 6), visualized with XGI \cite{landry2023xgi}. Blobs represent decision groups. Nodes are shown in white if part of a single decision group, blue if part of two, and orange if part of three.}
    \label{fig:polarized}
\end{figure}

We can explore an even more polarized population by having only zealots with strong opinions on all issues and with a more equal 60/40 split between types of agents.
In this scenario, all agents are internally coherent, but the two sub-populations have beliefs drawn from each of the two coherent solutions with opinions from truncated normal of mean $\pm 1$ and standard deviation $0.1$.
We then again vary the group size and overlap between discussion groups.
The results are shown in Fig.~\ref{fig:polarized}.

Our results show the power of networked governance and the importance of overlap between discussion groups.
Individuals who are part of the overlap have updated opinions that reflect the opinions of multiple other individuals with whom they have already discussed.
Consequently, we find that a large group (e.g., size 51) with a small overlap with previous groups (e.g., 5 to 14 in Fig.\ref{fig:zealots} and 1 to 3 in Fig.\ref{fig:polarized}) can reach almost the same or even stronger democratic satisfaction than random selection while also maximizing coherence.
Increasing overlap further might decrease satisfaction as fewer and fewer unique individuals get involved in the decision groups.
Importantly, only a small overlap is necessary to help a population reach a coherent set of decisions that respect the logical constraints.
We find a regime of \textit{effective governance} when we use large enough groups with small overlap with one another.
An example of the resulting governance hypergraph is shown in Fig.~\ref{fig:polarized}(panel F).

\begin{figure}
    \centering
    \includegraphics[width=0.7\linewidth]{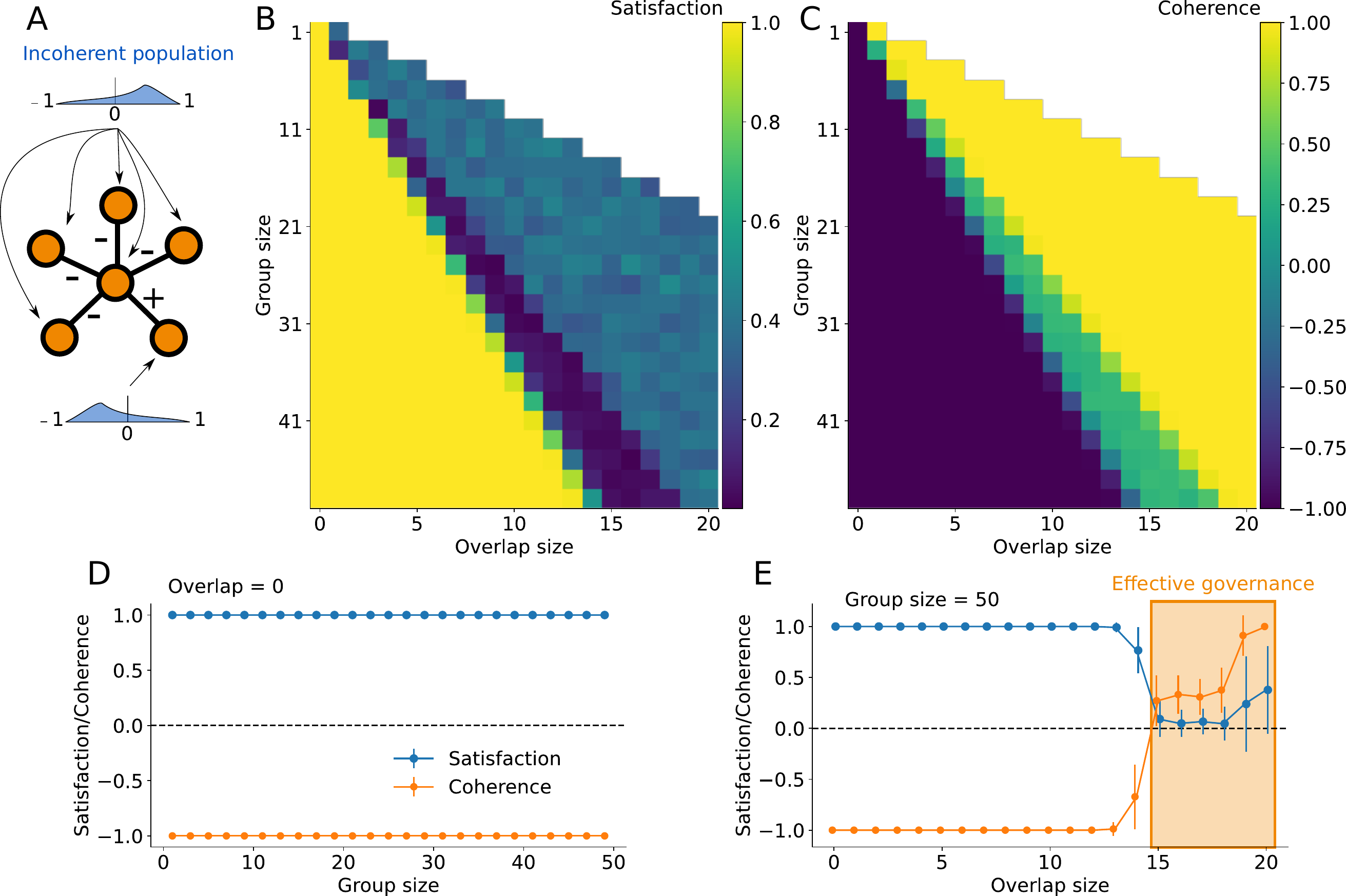}
    \caption{Governance model with an incoherent population. The simple decision network is illustrated in panel \textbf{A} and admits two complementary solutions.
    The population making the decision is an incoherent population with random independent opinions drawn uniformly from positive values with a probability of 0.6 and from negative values with a probability of 0.4 (standard deviation of 0.1).
    Increasing the overlap between decision groups can decrease satisfaction (panel \textbf{B}) but increase the coherence of the decisions (panel \textbf{C}), given that the population itself is homogeneous and incoherent (panel \textbf{D}).
    This suggests that the governance system can either be democratic \textit{or} coherent but not both.
    However, with significant overlap between large decision groups, we can dramatically increase coherence while maintaining some positive satisfaction.
    This is again the \textit{effective governance} region, highlighted in panel \textbf{E}.}
    \label{fig:incoherent}
\end{figure}

We also explore another problematic population, one in which all agents agree but have incoherent beliefs about what decisions the population should make.
Opinions for all agents are once again drawn from a normal distribution with a standard deviation of 0.1 and truncated to the $[-1,1]$ interval. 
Opinion means are distributed so that agents have, on average, a positive 0.6 opinion of the central decision, a weaker but still positive opinion (value of 0.4) of all decisions involved in exclusive clauses (``or'') clause with the central decision, and a strong negative -0.6 opinion of the decision with a positive (``and'') associative clause with the central decision.
The results of this experiment are shown in Fig.\ref{fig:incoherent}.
In principle, there is no good governance for such a population since there is a perfect one-to-one trade-off between democratic satisfaction and coherence of decisions.
Yet, we surprisingly find a regime where maximizing the overlap between decision groups can find a positive local maximum of both satisfaction and coherence.

\subsection{Increasingly hard problems}

As a second type of experiment, we assume that all agents in our population are independent greedy solvers of randomly generated satisfiability problems (or decision networks).
In this experimental design, we initialize the opinion of agents by randomly selecting a first positive opinion and snowballing our way through connected constraints, picking coherent solutions as we go, and setting other random positive opinions when there are no connected constraints.
With this process, every agent has their own random set of greedy (and close to maximally) coherent opinions.

\begin{figure}
    \centering
    \includegraphics[width=0.8\linewidth]{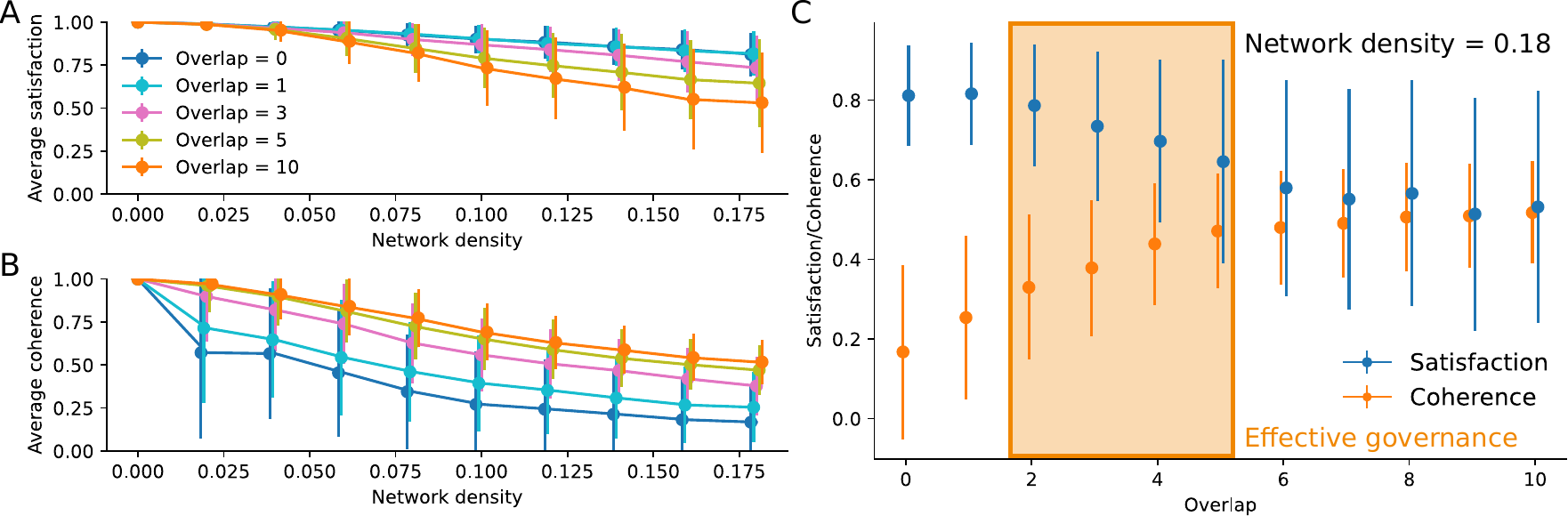}
    \caption{We use a population of 500 agents, each with their own greedy coherent opinions, about 20 decisions connected with random constraints of random signs based on a fixed network density.
    This leads to a population polarized around every exclusive (or negative) constraint.
    We fix the size of decision groups to 20 and test the ability of this population to govern as we vary overlap (curves) and density of random constraints.
    We shift the points slightly to avoid overlapping error bars.
    Without overlap, satisfaction is always maximized as we are simply sampling the greedy agents (panel \textbf{A}), but coherence falls dramatically as the density of constraints is increased (panel \textbf{B}).
    By tuning the overlap we can balance democratic satisfaction and coherence of the decisions (panel \textbf{C}).
    Especially at low but non-zero overlap, around 3, we find a large gain in coherence at a very low satisfaction cost (always within a standard deviation of the top-performing average).}
    \label{fig:random}
\end{figure}

We then vary the density of constraints and test the ability of a population of greedy solvers to find coherent decisions that also please a majority of agents.
In the limit of very sparse decision networks, where the number of constraints goes to zero, the decision problem is trivial.
However, there is a known transition in which the problem becomes harder as we randomly add constraints to the network \cite{varga2016order}.
We are therefore curious about the possibility of a networked governance system to improve the decision made by an ensemble of independent greedy solvers. 
Our results are shown in Fig.\ref{fig:random}.

We again find that effective governance can be reached with large enough groups and moderate overlap.
The key feature of our results is that, as we increase the overlap between decision groups, the coherence of decisions increases faster than demographic satisfaction decreases. 
For instance, at a network density of 0.18 and with group sizes of 20 agents, going from an overlap of zero to two decreases the average satisfaction by 3\% only while the average coherence increases by 57\%.

\section{Discussion}
\label{section:discussion}

We developed a model through which small subgroups within much larger populations can come together in a structured, networked way to efficiently make complicated decisions that balance the democratic opinions of the whole population and logical constraints between decisions.
At its core, the model is a mix between a hypergraph voter model and a satisfiability problem, where the satisfiability problem can influence the dynamics of the voter models and where the opinions of the voting population can complicate the satisfiability problem.

We investigated the behaviour of this model on toy and random satisfiability problems using different populations consisting of polarized or incoherent opinions and populations of greedy solvers. These are only a few small explorations of what is possible in our framework. Much more complex computational experiments could be explored in future work.
Despite their simplicity, our experiments consistently identify a regime of effective governance where mostly coherent decisions can be achieved while also respecting the democratic opinions of the larger populations on most issues.

From a top-down perspective, it is interesting to think of governments and other governance structures as solvers of ``democratic satisfiability problems'' where one must not only find a solution to logical constraints but do so while maximizing the proximity of the solution to the opinion distribution of a population. 
How should one solve these democratic satisfiability problems?
How likely is it that hard problems are made impossible by having an adversarial (e.g., polarized) population?
Some of these questions remain to be seen.

Perhaps more interestingly, we have shown the power of the bottom-up perspective.
A population of agents can be structured in an intelligent way to efficiently solve the satisfiability problem on their own by forming a network of discussions to reach some form of consensus.
Our model can, therefore, be used to explore governance practices and how decisions are shaped by the organizational network structure.
How large should the decision groups be?
How should participants be invited?
And how many bridges to other groups exist?
Other related questions should be explored in future work.
How important is it to view governance as an experimental process, revisiting decisions with different groups over time?
How sensitive to the nature of the discussion process are the solutions?
For example, what if some agents do not change their mind?
What if some agents occupy the discussion more than others?
Governance models might increase citizen participation in collective decision-making processes, amplifying both the strengths and weaknesses of distributed governance.
Indeed, some technical decisions may be better solved by individual experts, or even artificial intelligence.
And in some cases, individuals may not fully grasp their own potential contribution to group-level solutions \cite{meluso2023multidisciplinary}.
We could imagine including other individual dimensions in the model, such as expertise and confidence (which may or may not be correlated), to explore tasks with different requirements at the intersection of collective intelligence and governance. 
We would likely find that no single governance structure can efficiently solve decision problems of all kinds.


While full-scale implementation of such governance structures at a societal level is obviously a long way off, we foresee a strong possibility for two branches of empirical research that could accelerate progress in conjunction with further modeling like ours.
First, an observational branch could characterize governance structures in many smaller-scale organizations (corporations of various sizes and various divisions within them, universities and their schools and departments, non-governmental organizations, government administrative branches, professional societies, etc.).
It is possible that relatively efficient solutions tend to emerge on their own in a way that is adapted to the context.
Characterizing the constraints, types of decisions, network structures, and coherence/satisfaction of existing organizations, along with models such as ours to examine the theoretically optimal solutions, could provide important insight both on whether good structures emerge on their own and on what structures are best.
Second, an experimental branch of research could attempt to alter governance structures of small-scale organizations that are willing to participate, testing for improvements in satisfiability and coherence.
This could involve both minor changes to existing structures and wholesale reorganization.
Of course, satisfaction and coherence are not the only two criteria for appropriate decision-making in governance.
It is useful to think of the performance of a given governance structure along multiple axes, for example:
\begin{enumerate}
    \item As explored here, coherence, or the optimality of the decisions (i.e., minimizing constraint penalties).
    \item As also explored here: satisfaction (i.e., minimizing democratic penalties).
    \item Time or number of operations required to make specific time-sensitive decisions (e.g., response to military attack).
    \item Time, effort, or costs required for the entire process of decision-making (i.e., feasibility of the process).
    \item Compliance with decisions made and public perception about the decision process itself.
    \item Robustness of the governance structure to outside influences, misinformed individuals, random failures, or targeted attacks.
\end{enumerate}
Other important dimensions could prove critical but are harder to measure or include in our current framework.
One particularly subtle benefit of governance structures is innovation, since governance groups operate with discussions, and brainstorming can lead to novel solutions not initially available.

Despite these future empirical and theoretical challenges, our framework shows that taking a complex system and modeling-focused perspective on governance can help us understand decentralized decision-making as a growing hypergraph.
This approach to governance is more efficient than direct democracy in the sense that it requires fewer votes, but it also achieves better solutions if the members of groups, or hyperedges in the discussion network, update their opinions as part of the decision-making process.
Decentralized governance, as conceptualized in this framework, is therefore well-suited for societal crises that require a fast consensus of experts and stakeholders, such as epidemics, emerging technologies, and climate change.
However, future work---especially empirical work---will be required to connect our abstract framework to real-world problems.

\section*{Data availability}
No data were created or analyzed in this study.

\section*{Code availability}
A repository for this project, containing codes to reproduce all results, is available online\cite{software}.

\section*{Acknowledgements}
The authors acknowledge Dominique Gravel for discussions. This work was supported by the Fonds de recherche du Qu\'ebec through an AUDACE award (S.B., C.C., and A.C.), by The Alfred P. Sloan Foundation (L.H.-D., J.L. and J.S.-O.), by Google Open Source through the Open-Source Complex Ecosystems And Networks (OCEAN) project (L.H.-D., J.L., J.S.-O. and J.-G.Y.), by the National Science Foundation awards OIA-2019470 (L.H.-D.), OIA-2242829 (J.L.), and SES-2419733 (L.H.-D. and J.-G.Y.), by the National Institutes of Health 1P20 GM125498-01 Centers of Biomedical Research Excellence Award (L.H.-D., N.W.L., and J.-G.Y.), and by the University of Virginia Prominence-to-Preeminence (P2PE) STEM Targeted Initiatives Fund SIF176A Contagion Science (N.W.L.). Our theoretical model was inspired by the field research achieved by the Acclimatons-nous project (\href{www.acclimatons-nous.org}{www.acclimatons-nous.org}).

\section*{Author contributions}
L.H.-D., J.L., J.S.-O., S.B., C.C., and A.A.C. conceived the model. L.H.-D., N.W.L. J.L., J.S.-O., and J.-G.Y. implemented the model. L.H.-D. designed and ran the experiments. M.-E.C.-M., S.B., C.C., and A.A.C. conceptualized the study. S.B., C.C., and A.A.C. managed the project. All authors wrote and reviewed the manuscript.

\section*{Competing interests}
The authors declare no competing interests.


\begin{thebibliography}{10}
\urlstyle{rm}
\expandafter\ifx\csname url\endcsname\relax
  \def\url#1{\texttt{#1}}\fi
\expandafter\ifx\csname urlprefix\endcsname\relax\def\urlprefix{URL }\fi
\expandafter\ifx\csname doiprefix\endcsname\relax\def\doiprefix{DOI: }\fi
\providecommand{\bibinfo}[2]{#2}
\providecommand{\eprint}[2][]{\url{#2}}

\bibitem{wolfers2002voters}
\bibinfo{author}{Wolfers, J.} \emph{et~al.}
\newblock \emph{\bibinfo{title}{Are voters rational? Evidence from
  gubernatorial elections}} (\bibinfo{publisher}{Citeseer},
  \bibinfo{year}{2002}).

\bibitem{smidt2017polarization}
\bibinfo{author}{Smidt, C.~D.}
\newblock \bibinfo{journal}{\bibinfo{title}{Polarization and the decline of the
  american floating voter}}.
\newblock {\emph{\JournalTitle{American Journal of Political Science}}}
  \textbf{\bibinfo{volume}{61}}, \bibinfo{pages}{365--381}
  (\bibinfo{year}{2017}).

\bibitem{spencer1998mathematical}
\bibinfo{author}{Spencer, C.~S.}
\newblock \bibinfo{journal}{\bibinfo{title}{A mathematical model of primary
  state formation}}.
\newblock {\emph{\JournalTitle{Cultural Dynamics}}}
  \textbf{\bibinfo{volume}{10}}, \bibinfo{pages}{5--20} (\bibinfo{year}{1998}).

\bibitem{laver1998models}
\bibinfo{author}{Laver, M.}
\newblock \bibinfo{journal}{\bibinfo{title}{Models of government formation}}.
\newblock {\emph{\JournalTitle{Annual Review of Political Science}}}
  \textbf{\bibinfo{volume}{1}}, \bibinfo{pages}{1--25} (\bibinfo{year}{1998}).

\bibitem{kerchove2019pyramide}
\bibinfo{author}{Van~de Kerchove, M.} \& \bibinfo{author}{Ost, F.}
\newblock \emph{\bibinfo{title}{De la pyramide au r{\'e}seau?: pour une
  th{\'e}orie dialectique du droit}} (\bibinfo{publisher}{Presses de
  l’Universit{\'e} Saint-Louis}, \bibinfo{year}{2019}).

\bibitem{choquette2021barrages}
\bibinfo{author}{Choquette, C.} \emph{et~al.}
\newblock \bibinfo{journal}{\bibinfo{title}{L'adaptation de la gestion des
  barrages aux changements climatiques}}.
\newblock {\emph{\JournalTitle{Le Climatoscope}}} \textbf{\bibinfo{volume}{3}},
  \bibinfo{pages}{95--101} (\bibinfo{year}{2021}).

\bibitem{commaille2019metamorphoses}
\bibinfo{author}{Commaille, J.} \& \bibinfo{author}{Jobert, B.}
\newblock \emph{\bibinfo{title}{Les m{\'e}tamorphoses de la r{\'e}gulation
  politique}} (\bibinfo{publisher}{LGDJ}, \bibinfo{year}{2019}).

\bibitem{rocher2019pluralite}
\bibinfo{author}{Rocher, G.}
\newblock \bibinfo{journal}{\bibinfo{title}{La pluralit{\'e} des ordres
  juridiques}}.
\newblock {\emph{\JournalTitle{Rev. Gen.}}} \textbf{\bibinfo{volume}{49}},
  \bibinfo{pages}{443} (\bibinfo{year}{2019}).

\bibitem{keast2022network}
\bibinfo{author}{Keast, R.}
\newblock \bibinfo{title}{Network governance}.
\newblock In \emph{\bibinfo{booktitle}{Handbook on theories of governance}},
  \bibinfo{pages}{485--496} (\bibinfo{publisher}{Edward Elgar Publishing},
  \bibinfo{year}{2022}).

\bibitem{fischer2012participatory}
\bibinfo{author}{Fischer, F.}
\newblock \bibinfo{journal}{\bibinfo{title}{Participatory governance: From
  theory to practice}}.
\newblock {\emph{\JournalTitle{Oxford Handbooks Online}}}
  (\bibinfo{year}{2012}).

\bibitem{couture2023expertise}
\bibinfo{author}{Couture-M{\'e}nard, M.-{\`E}.} \emph{et~al.}
\newblock \bibinfo{journal}{\bibinfo{title}{L'expertise et l'information dans
  la gouvernance de la crise sanitaire au qu{\'e}bec}}.
\newblock {\emph{\JournalTitle{Revue g{\'e}n{\'e}rale de droit}}}
  \textbf{\bibinfo{volume}{53}}, \bibinfo{pages}{133--175}
  (\bibinfo{year}{2023}).

\bibitem{couture2022gouvernance}
\bibinfo{author}{Couture-M{\'e}nard, M.-{\`E}.} \emph{et~al.}
\newblock \bibinfo{journal}{\bibinfo{title}{La gouvernance, nouvel
  environnement de l'exercice des pouvoirs publics en temps de crise sanitaire:
  l'exemple de la r{\'e}gulation des masques}}.
\newblock {\emph{\JournalTitle{{\'E}thique publique}}}
  \textbf{\bibinfo{volume}{24}} (\bibinfo{year}{2022}).

\bibitem{axelrod1997dissemination}
\bibinfo{author}{Axelrod, R.}
\newblock \bibinfo{journal}{\bibinfo{title}{The dissemination of culture: A
  model with local convergence and global polarization}}.
\newblock {\emph{\JournalTitle{Journal of conflict resolution}}}
  \textbf{\bibinfo{volume}{41}}, \bibinfo{pages}{203--226}
  (\bibinfo{year}{1997}).

\bibitem{rodriguez2007smartocracy}
\bibinfo{author}{Rodriguez, M.~A.} \emph{et~al.}
\newblock \bibinfo{title}{Smartocracy: Social networks for collective decision
  making}.
\newblock In \emph{\bibinfo{booktitle}{2007 40th Annual Hawaii International
  Conference on System Sciences (HICSS'07)}}, \bibinfo{pages}{90--90}
  (\bibinfo{organization}{IEEE}, \bibinfo{year}{2007}).

\bibitem{cohen2012physiological}
\bibinfo{author}{Cohen, A.~A.}, \bibinfo{author}{Martin, L.~B.},
  \bibinfo{author}{Wingfield, J.~C.}, \bibinfo{author}{McWilliams, S.~R.} \&
  \bibinfo{author}{Dunne, J.~A.}
\newblock \bibinfo{journal}{\bibinfo{title}{Physiological regulatory networks:
  ecological roles and evolutionary constraints}}.
\newblock {\emph{\JournalTitle{Trends in Ecology \& Evolution}}}
  \textbf{\bibinfo{volume}{27}}, \bibinfo{pages}{428--435}
  (\bibinfo{year}{2012}).

\bibitem{cohen2022complex}
\bibinfo{author}{Cohen, A.~A.} \emph{et~al.}
\newblock \bibinfo{journal}{\bibinfo{title}{A complex systems approach to aging
  biology}}.
\newblock {\emph{\JournalTitle{Nature Aging}}} \textbf{\bibinfo{volume}{2}},
  \bibinfo{pages}{580--591} (\bibinfo{year}{2022}).

\bibitem{bernatchez2021liber}
\bibinfo{author}{Bernatchez, S.} \& \bibinfo{author}{Couture-M{\'e}nard,
  M.-{\`E}.}
\newblock \bibinfo{title}{Pour ce liber amicorum, le droit de la gouvernance
  pr{\'e}sent{\'e} {\`a} l'amicus kouri {\`a} partir du cas de la
  r{\'e}gulation de la pand{\'e}mie de {COVID-19}}.
\newblock In \emph{\bibinfo{booktitle}{L'humain au c\oe{}ur du droit}},
  \bibinfo{pages}{447--477} (\bibinfo{publisher}{\'Editions Yvon Blais},
  \bibinfo{year}{2021}).

\bibitem{wang2007satisfiability}
\bibinfo{author}{Wang, Q.} \& \bibinfo{author}{Li, N.}
\newblock \bibinfo{title}{Satisfiability and resiliency in workflow systems}.
\newblock In \emph{\bibinfo{booktitle}{European Symposium on Research in
  Computer Security}}, \bibinfo{pages}{90--105}
  (\bibinfo{organization}{Springer}, \bibinfo{year}{2007}).

\bibitem{mohammed2022pandemic}
\bibinfo{author}{Boughrous, M.}, \bibinfo{author}{El~Bakkali, H.} \&
  \bibinfo{author}{El~Kandoussi, A.}
\newblock \bibinfo{title}{The pandemic impact on organizations security and
  resiliency: The workflow satisfiability problem}.
\newblock In \emph{\bibinfo{booktitle}{Hybrid Intelligent Systems: 21st
  International Conference on Hybrid Intelligent Systems (HIS 2021), December
  14-16, 2021}}, vol. \bibinfo{volume}{420}, \bibinfo{pages}{321}
  (\bibinfo{organization}{Springer Nature}, \bibinfo{year}{2022}).

\bibitem{moore2011nature}
\bibinfo{author}{Moore, C.} \& \bibinfo{author}{Mertens, S.}
\newblock \emph{\bibinfo{title}{The nature of computation}}
  (\bibinfo{publisher}{Oxford University Press}, \bibinfo{year}{2011}).

\bibitem{garey1979computers}
\bibinfo{author}{Garey, M.~R.} \& \bibinfo{author}{Johnson, D.~S.}
\newblock \emph{\bibinfo{title}{Computers and intractability}}, vol.
  \bibinfo{volume}{174} (\bibinfo{publisher}{freeman San Francisco},
  \bibinfo{year}{1979}).

\bibitem{hickok2022bounded}
\bibinfo{author}{Hickok, A.}, \bibinfo{author}{Kureh, Y.},
  \bibinfo{author}{Brooks, H.~Z.}, \bibinfo{author}{Feng, M.} \&
  \bibinfo{author}{Porter, M.~A.}
\newblock \bibinfo{journal}{\bibinfo{title}{A bounded-confidence model of
  opinion dynamics on hypergraphs}}.
\newblock {\emph{\JournalTitle{SIAM Journal on Applied Dynamical Systems}}}
  \textbf{\bibinfo{volume}{21}}, \bibinfo{pages}{1--32} (\bibinfo{year}{2022}).

\bibitem{battiston2020networks}
\bibinfo{author}{Battiston, F.} \emph{et~al.}
\newblock \bibinfo{journal}{\bibinfo{title}{Networks beyond pairwise
  interactions: structure and dynamics}}.
\newblock {\emph{\JournalTitle{Physics Reports}}}
  \textbf{\bibinfo{volume}{874}}, \bibinfo{pages}{1--92}
  (\bibinfo{year}{2020}).

\bibitem{aksoy2020hypernetwork}
\bibinfo{author}{Aksoy, S.~G.}, \bibinfo{author}{Joslyn, C.},
  \bibinfo{author}{Marrero, C.~O.}, \bibinfo{author}{Praggastis, B.} \&
  \bibinfo{author}{Purvine, E.}
\newblock \bibinfo{journal}{\bibinfo{title}{Hypernetwork science via high-order
  hypergraph walks}}.
\newblock {\emph{\JournalTitle{EPJ Data Science}}}
  \textbf{\bibinfo{volume}{9}}, \bibinfo{pages}{16} (\bibinfo{year}{2020}).

\bibitem{overgoor2019choosing}
\bibinfo{author}{Overgoor, J.}, \bibinfo{author}{Benson, A.} \&
  \bibinfo{author}{Ugander, J.}
\newblock \bibinfo{title}{Choosing to grow a graph: Modeling network formation
  as discrete choice}.
\newblock In \emph{\bibinfo{booktitle}{The World Wide Web Conference}},
  \bibinfo{pages}{1409--1420} (\bibinfo{year}{2019}).

\bibitem{clifford1973model}
\bibinfo{author}{Clifford, P.} \& \bibinfo{author}{Sudbury, A.}
\newblock \bibinfo{journal}{\bibinfo{title}{A model for spatial conflict}}.
\newblock {\emph{\JournalTitle{Biometrika}}} \textbf{\bibinfo{volume}{60}},
  \bibinfo{pages}{581--588} (\bibinfo{year}{1973}).

\bibitem{holley1975ergodic}
\bibinfo{author}{Holley, R.~A.} \& \bibinfo{author}{Liggett, T.~M.}
\newblock \bibinfo{journal}{\bibinfo{title}{Ergodic theorems for weakly
  interacting infinite systems and the voter model}}.
\newblock {\emph{\JournalTitle{The annals of probability}}}
  \bibinfo{pages}{643--663} (\bibinfo{year}{1975}).

\bibitem{redner2019reality}
\bibinfo{author}{Redner, S.}
\newblock \bibinfo{journal}{\bibinfo{title}{Reality-inspired voter models: A
  mini-review}}.
\newblock {\emph{\JournalTitle{Comptes Rendus Physique}}}
  \textbf{\bibinfo{volume}{20}}, \bibinfo{pages}{275--292}
  (\bibinfo{year}{2019}).

\bibitem{castellano2000nonequilibrium}
\bibinfo{author}{Castellano, C.}, \bibinfo{author}{Marsili, M.} \&
  \bibinfo{author}{Vespignani, A.}
\newblock \bibinfo{journal}{\bibinfo{title}{Nonequilibrium phase transition in
  a model for social influence}}.
\newblock {\emph{\JournalTitle{Physical Review Letters}}}
  \textbf{\bibinfo{volume}{85}}, \bibinfo{pages}{3536} (\bibinfo{year}{2000}).

\bibitem{sood2005voter}
\bibinfo{author}{Sood, V.} \& \bibinfo{author}{Redner, S.}
\newblock \bibinfo{journal}{\bibinfo{title}{Voter model on heterogeneous
  graphs}}.
\newblock {\emph{\JournalTitle{Physical Review Letters}}}
  \textbf{\bibinfo{volume}{94}}, \bibinfo{pages}{178701}
  (\bibinfo{year}{2005}).

\bibitem{horstmeyer2020adaptive}
\bibinfo{author}{Horstmeyer, L.} \& \bibinfo{author}{Kuehn, C.}
\newblock \bibinfo{journal}{\bibinfo{title}{Adaptive voter model on simplicial
  complexes}}.
\newblock {\emph{\JournalTitle{Physical Review E}}}
  \textbf{\bibinfo{volume}{101}}, \bibinfo{pages}{022305}
  (\bibinfo{year}{2020}).

\bibitem{masuda2010heterogeneous}
\bibinfo{author}{Masuda, N.}, \bibinfo{author}{Gibert, N.} \&
  \bibinfo{author}{Redner, S.}
\newblock \bibinfo{journal}{\bibinfo{title}{Heterogeneous voter models}}.
\newblock {\emph{\JournalTitle{Physical Review E}}}
  \textbf{\bibinfo{volume}{82}}, \bibinfo{pages}{010103}
  (\bibinfo{year}{2010}).

\bibitem{mobilia2007role}
\bibinfo{author}{Mobilia, M.}, \bibinfo{author}{Petersen, A.} \&
  \bibinfo{author}{Redner, S.}
\newblock \bibinfo{journal}{\bibinfo{title}{On the role of zealotry in the
  voter model}}.
\newblock {\emph{\JournalTitle{Journal of Statistical Mechanics: Theory and
  Experiment}}} \textbf{\bibinfo{volume}{2007}}, \bibinfo{pages}{P08029}
  (\bibinfo{year}{2007}).

\bibitem{castellano2009nonlinear}
\bibinfo{author}{Castellano, C.}, \bibinfo{author}{Mu{\~n}oz, M.~A.} \&
  \bibinfo{author}{Pastor-Satorras, R.}
\newblock \bibinfo{journal}{\bibinfo{title}{Nonlinear q-voter model}}.
\newblock {\emph{\JournalTitle{Physical Review E}}}
  \textbf{\bibinfo{volume}{80}}, \bibinfo{pages}{041129}
  (\bibinfo{year}{2009}).

\bibitem{vazquez2004ultimate}
\bibinfo{author}{Vazquez, F.} \& \bibinfo{author}{Redner, S.}
\newblock \bibinfo{journal}{\bibinfo{title}{Ultimate fate of constrained
  voters}}.
\newblock {\emph{\JournalTitle{Journal of Physics A: Mathematical and
  General}}} \textbf{\bibinfo{volume}{37}}, \bibinfo{pages}{8479}
  (\bibinfo{year}{2004}).

\bibitem{hebert2011structural}
\bibinfo{author}{H{\'e}bert-Dufresne, L.}, \bibinfo{author}{Allard, A.},
  \bibinfo{author}{Marceau, V.}, \bibinfo{author}{No{\"e}l, P.-A.} \&
  \bibinfo{author}{Dub{\'e}, L.~J.}
\newblock \bibinfo{journal}{\bibinfo{title}{Structural preferential attachment:
  Network organization beyond the link}}.
\newblock {\emph{\JournalTitle{Physical Review Letters}}}
  \textbf{\bibinfo{volume}{107}}, \bibinfo{pages}{158702}
  (\bibinfo{year}{2011}).

\bibitem{landry2023xgi}
\bibinfo{author}{Landry, N.~W.} \emph{et~al.}
\newblock \bibinfo{journal}{\bibinfo{title}{{XGI: A Python package for
  higher-order interaction networks}}}.
\newblock {\emph{\JournalTitle{Journal of Open Source Software}}}
  \textbf{\bibinfo{volume}{8}}, \bibinfo{pages}{5162},
  \doiprefix\url{10.21105/joss.05162} (\bibinfo{year}{2023}).

\bibitem{varga2016order}
\bibinfo{author}{Varga, M.}, \bibinfo{author}{Sumi, R.},
  \bibinfo{author}{Toroczkai, Z.} \& \bibinfo{author}{Ercsey-Ravasz, M.}
\newblock \bibinfo{journal}{\bibinfo{title}{Order-to-chaos transition in the
  hardness of random boolean satisfiability problems}}.
\newblock {\emph{\JournalTitle{Physical Review E}}}
  \textbf{\bibinfo{volume}{93}}, \bibinfo{pages}{052211}
  (\bibinfo{year}{2016}).

\bibitem{meluso2023multidisciplinary}
\bibinfo{author}{Meluso, J.} \& \bibinfo{author}{H{\'e}bert-Dufresne, L.}
\newblock \bibinfo{journal}{\bibinfo{title}{Multidisciplinary learning through
  collective performance favors decentralization}}.
\newblock {\emph{\JournalTitle{Proceedings of the National Academy of
  Sciences}}} \textbf{\bibinfo{volume}{120}}, \bibinfo{pages}{e2303568120}
  (\bibinfo{year}{2023}).

\bibitem{software}
\bibinfo{author}{Landry, N.}, \bibinfo{author}{H\'ebert-Dufresne, L.},
  \bibinfo{author}{Lovato, J.} \& \bibinfo{author}{St-Onge, J.}
\newblock \bibinfo{journal}{\bibinfo{title}{Governance repository v0.0}}.
\newblock {\emph{\JournalTitle{Zenodo}}}
  \doiprefix\url{10.5281/zenodo.14257343} (\bibinfo{year}{2024}).

\end{thebibliography}
\end{document}